\documentclass[twocolumn,showpacs,amssymb]{revtex4}
\usepackage{graphicx}
\usepackage{txfonts}
\usepackage{natbib}

\begin{document}

\title{Estimating the uncorrelated dark energy evolution in the Planck era}

\author{F. Y. Wang$^{1,2\ast}$ and Z. G. Dai$^{1,2\dagger}$}

\affiliation{$^1$School of Astronomy and Space Science, Nanjing University, Nanjing 210093, China\\
$^2$Key laboratory of Modern Astronomy and Astrophysics (Nanjing
University), Ministry of Education, Nanjing 210093, China \\
$^\ast$fayinwang@nju.edu.cn, $^\dagger$dzg@nju.edu.cn}

\begin{abstract}
The equation of state (EOS), $w(z)$, is the most important parameter
of dark energy. We reconstruct the evolution of this EOS in a
model-independent way using the latest cosmic microwave background
(CMB) data from Planck and other observations, such as type Ia
supernovae (SNe Ia), the baryonic acoustic oscillation measurements
(SDSS, 6dF, BOSS, and WiggleZ), and the Hubble parameter value
$H(z)$. The results show that the EOS is consistent with the
cosmological constant at the $2\sigma$ confidence level, not
preferring a dynamical dark energy. The uncorrelated EOS of dark
energy constraints from Planck CMB data are much tighter than those
from the WMAP 9-year CMB data.
\end{abstract}

\date{\today}
\maketitle

\section{Introduction}

The acceleration of our Universe was first discovered by observing
type Ia supernovae (SNe Ia) \cite{perl,riess}. This unexpected
discovery has been confirmed by the observations on cosmic microwave
background, large scale structure, weak gravitational lensing, and
Hubble parameter. Within the framework of general relativity, the
accelerating expansion is attributed to mysterious dark energy,
which has an equation of state (EOS), $w=p/\rho$, where $P$ and
$\rho$ are the pressure and energy density respectively. In addition
to the cosmological constant, other many dark energy models have
been suggested (for reviews see \cite{Weinberg12,Li11}).

EOS $w$ is the most important parameter that describes the
properties of dark energy. Whether and how it evolves with time is
crucial for revealing the physics of dark energy. The evolution of
EOS can be reconstructed from data using either parametric or
nonparametric methods \cite{Sahni}. Simple parametric forms of
$w(z)$ have been proposed for studying the evolution of dark energy,
such as $ w(z) = w_0 + w_1z$ \cite{Cooray99} and $w(z) = w_0 +
w_1z/(1+z)$ \cite{Chevallier01, Linder03}. Reconstructing the EOS of
dark energy in a parametric method has been widely explored
\cite{para}. However, a simplified dark energy parameterization is
equivalent to a strong and not necessarily theoretically justified
prior on the nature of dark energy \cite{Riess07}. In order to avoid
this shortcoming, the nonparametric approach was proposed
\cite{Huterer03,Huterer05,nonpara,Holsclaw10,gauss}. The procedure
is to bin $w$ in $z$, and fit the $w$ in each bin to data. This
method assumes that $w(z)$ is constant in each bin and has been
widely used in the literature. The procedure is as follows: first
performing a Fisher forecast to determine the eigenmodes of the
covariance matrix, and then fitting the constrained modes to
observational data. The Fisher matrix methodology has been used in
predicting the ability of future experiments to constrain particular
parameters of interest \cite{Tegmark}. The Joint Dark Energy Mission
team proposed that this method can constrain binned 36 EOS
parameters with future data \cite{Albrecht}. Another method has also
been proposed, which puts prior directly on the space of $w(z)$
functions \cite{space}. But the functions may be arbitrary. The
Gaussian process is another method of dealing with stochastic
variables \cite{Shafieloo,Holsclaw10}.

Using the nonparametric methods, two distinctive evolution of $w(z)$
have been derived, one is consistent with the cosmological constant
\cite{Riess07,Huterer05,Holsclaw10,Sullivan07}, and the other
supports a dynamical dark energy model \cite{Qi09,Zhao12}. Zhao et
al. found that a dynamical dark energy model which evolves from
$w<-1$ at $z\sim0.25$ to $z>-1$ at higher redshift is favored using
the WMAP 7-year data and other cosmological observations
\cite{Zhao12}. So the evolutional behavior of $w(z)$ is
controversial. More recently, the Planck team has released the first
cosmological results based on the measurement of the CMB temperature
and lensing-potential power spectra, which prefer a low Hubble
constant and high matter density. In this paper, we estimate the
evolution of $w(z)$ using the latest cosmological observations
including the Planck data.

\section{Method and results}

We use the nonparametric technique to constrain the uncorrelated EOS
of dark energy $w(z)$ \cite{Huterer05}. This is a modification
version of the method proposed by \cite{Huterer03}. Theoretically,
the luminosity distance $d_L(z)$ is given by
\begin{equation}
  \label{eq:dlz_theoretical}
  d_L(z)=(1+z)\frac{c}{H_0}\times
  \left\{
    \begin{array}{ll}
      \frac{1}{\sqrt{|\Omega_k|}}
      \sinh\left(
        \sqrt{|\Omega_k|} \int_0^z\frac{\mathrm{d}\tilde{z}}{E(\tilde{z})}
      \right) & \textrm{if } \Omega_k>0
      \\
      \int_0^z\frac{\mathrm{d}\tilde{z}}{E(\tilde{z})}
      & \textrm{if } \Omega_k=0
      \\
      \frac{1}{\sqrt{|\Omega_k|}}
      \sin\left(
        \sqrt{|\Omega_k|} \int_0^z\frac{\mathrm{d}\tilde{z}}{E(\tilde{z})}
      \right) & \textrm{if } \Omega_k<0
    \end{array}
  \right.
  ,
\end{equation}
where $E(z)=[\Omega_m (1+z)^3+\Omega_x f(z) + \Omega_k
(1+z)^2]^{1/2}$, $\Omega_m+\Omega_x+\Omega_k=1$, and
\begin{eqnarray}
  \label{eq:fz}
  f(z)=\exp \left[
    3\int_0^z\frac{1+w(\tilde{z})}{1+\tilde{z}}\mathrm{d}\tilde{z}
  \right]
  .
\end{eqnarray}
The function $f(z)$ is related to the parameterization of dark
energy. If the EOS is considered to be piecewise constant in
redshift, then $f(z)$ can be described as~\cite{Sullivan07}
\begin{equation}
  \label{eq:fzbinned}
  f(z_{n-1}<z \le z_n)=
  (1+z)^{3(1+w_n)}\prod_{i=0}^{n-1}(1+z_i)^{3(w_i-w_{i+1})},
\end{equation}
where $w_i$ is the EOS parameter in the $i^{\mathrm{th}}$ redshift
bin defined by an upper boundary at $z_i$, and the zeroth bin is
defined as $z_0=0$. We choose 10 bins in our analysis, e.g.
$z_i=0.1,0.2,0.3,0.4,0.5,0.6,0.7,0.8,0.9,1.4$. Previous studies only
used less than five redshift bins. We fix $w=-1$ at $z>1.4$. This
assumption does not affect the derived results
\cite{Sullivan07,Qi09,Zhao12}. The number of bins and the range are
chosen to be large enough so that the derived results are stable to
these choices.

The luminosity distance depends on the integration of the behavior
of the dark energy over redshift, so the EOS parameters $w_i$ are
correlated. The covariance matrix,
\begin{equation}
  \label{eq:cov_matrix}
  \mathbf{C}
  =\langle \mathbf{w} \mathbf{w}^{\mathrm{T}} \rangle
  - \langle \mathbf{w} \rangle
  \langle \mathbf{w}^{\mathrm{T}} \rangle,
\end{equation}
is not diagonal. In the above equation, the $\mathbf{w}$ is a vector
with components $w_i$ and the average is calculated by letting
$\mathbf{w}$ run over the Markov chain. We can obtain a set of
decorrelated parameters $\widetilde{w}_i$ through diagonalization of
the covariance matrix by choosing an appropriate transformation
\begin{equation}
  \label{eq:transformation}
  \widetilde{\mathbf{w}}=\mathbf{T} \mathbf{w}.
\end{equation}
We use the transformation proposed by~\cite{Huterer05}. First the
Fisher matrix is
\begin{equation}
  \label{eq:fisher_matrix}
  \mathbf{F}\equiv\mathbf{C}^{-1}
  =\mathbf{O}^{\mathrm{T}}\mathbf{\Lambda}\mathbf{O}
  ,
\end{equation}
and the transformation matrix $\mathbf{T}$ is given by
\begin{equation}
  \label{eq:transf_matrix1}
  \mathbf{T}=\mathbf{O}^{\mathrm{T}}
  \mathbf{\Lambda}^{\frac{1}{2}}\mathbf{O}
  .
\end{equation}
The advantage of this transformation is that the weights are
positive almost everywhere and localized in redshift fairly well, so
the uncorrelated EOS parameters $\widetilde{w}_i$ are easy to
interpret~\cite{Huterer05}.

\begin{figure}
\includegraphics[width=0.5\textwidth]{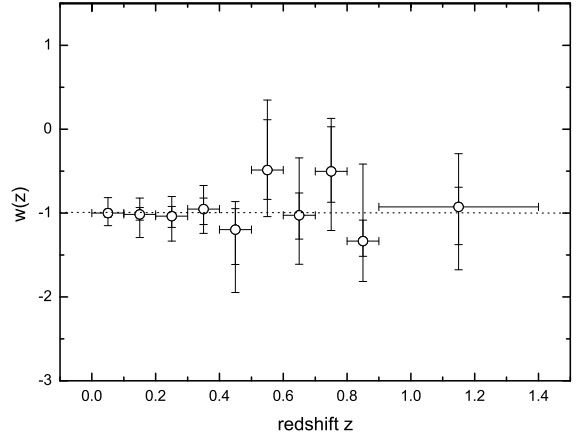}
\caption{\label{Fig1} Estimation of the uncorrelated dark energy EOS
parameters ($w_1,w_2,...,w_{10}$) at different redshift bins from
SNe Ia+BAO+Planck+H(z) data. The open points show the best fit
value. The error bars are $1\sigma$ and $2\sigma$ confidence levels.
The dotted line shows the cosmological constant.}
\end{figure}

\begin{figure}
\includegraphics[width=0.5\textwidth]{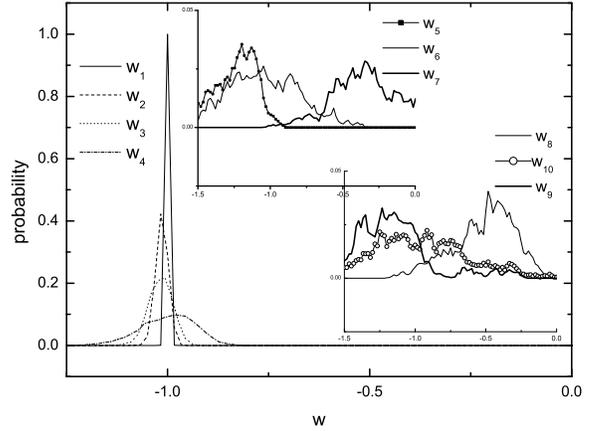}
\caption{\label{Fig2} The probability distributions of uncorrelated
dark energy EOS in each bins ($w_1,w_2,...,w_{10}$). The EOS is
tightly constrained at the first four bins. For other bins, the EOS
is poorly constrained. More high-redshift data is required to reduce
the error of EOS.}
\end{figure}

We apply the above method to a joint data set of the latest
observations including SNe Ia, CMB from Planck and WMAP
polarization, large scale structure and the Hubble parameter
measurement. We use the Union 2.1 SNe Ia sample \cite{Suzuki12}. We
use the distance priors \cite{Wang13} from Planck's first data
release \cite{PlanckI} and WMAP9 polarization \cite{Bennett12}, such
as CMB shift parameters $R=1.7407\pm 0.0094$, $l_a=301.57\pm 0.18$,
and $\omega_b\equiv \Omega_bh^2=0.02228\pm0.0003$. The distance
priors capture most of the information of the CMB data for dark
energy properties, and the constraints derived from the distance
prior are almost the same as those derived from the full analysis
\cite{prior}. For the baryonic acoustic oscillations (BAO) data, we
use the SDSS DR7 BAO measurements at effective redshifts $z=0.2$ and
$z=0.35$ \cite{Percival10}, the WiggleZ measurements at $z=0.44,
0.60, 0.73$ \cite{Blake11}, the BOSS DR9 measurement at $z=0.57$
\cite{Anderson13}, and the 6dF Galaxy Survey measurement at $z=0.1$
\cite{Beutler11}. The distance ratio vector of BAO is
\begin{eqnarray}
\hspace{-.5cm}{\bf{P}}_{\rm{BAO}}^{obs} &=& \left(\begin{array}{c}
{ d_{0.1}} \\
{d_{0.35}^{-1}}\\
{d_{0.57}^{-1}}\\
{ d_{0.44}} \\
{ d_{0.60}} \\
{ d_{0.73}} \\
\end{array}
  \right)=
  \left(\begin{array}{c}
0.336 \\
8.88\\
13.67\\
0.0916\\
0.0726\\
0.0592\\
\end{array}
  \right).
 \end{eqnarray}
The corresponding inverse covariance matrix is
\begin{eqnarray}
\hspace{-.75cm} {\bf
C_{\mathrm{BAO}}}^{-1}=\left(\begin{array}{rrrrrr}
4444.4 & 0 & 0 & 0 & 0 & 0 \\
0 & 34.602 & 0 & 0 & 0 & 0 \\
0 & 0 & 20.661157 & 0 & 0 & 0 \\
0 & 0 & 0 & 24532.1  & -25137.7 & 12099.1 \\
0 & 0 & 0 & -25137.7 & 134598.4 & -64783.9 \\
0 & 0 & 0 & 12099.1 & -64783.9 & 128837.6 \\
\end{array}
\right).
\end{eqnarray}
We also use the 28 independent measurements of the Hubble parameter
between redshifts $0.07\leq z\leq2.3$ compiled in \cite{Farooq13}.
There are 12 cosmological parameters in our analysis, such as
$\Omega_M, H_0, w_1,...,w_{10}$. We take $\Omega_M$ as a free
parameter in our calculation. According to the observations of
Planck, a flat cosmology is assumed.  We also adopt $\chi^2$
statistic to estimate parameters. The likelihood function
$\mathcal{L}$ is then proportional to $\exp\left(-\chi^2/2\right)$,
which produces the posterior probability when multiplied by the
prior probability. According to the posterior probability derived in
this way, Markov chains are generated through the Monte-Carlo
algorithm to study the statistical properties of the parameters. We
marginalize over the Hubble constant $73.8\pm 2.4$ $\rm
km~s^{-1}~Mpc^{-1}$ from Cepheid variables \cite{Riess11}.

Figure 1 shows the estimation of the uncorrelated dark energy EOS
parameters at different redshift bins. The dark energy EOS is
consistent with the cosmological constant at the $2\sigma$
confidence level. The EOS shows a marginal oscillation feature (from
$w<-1$ to $w>-1$) around redshift $z\sim 0.5$, but from $w>-1$ to
$w<-1$ at $z\sim0.8$. The reconstructed EOS is consistent with the
cosmological constant in all redshift bins at the $1 \sigma$
confidence level except for these three bins. In order to determine
the need for complex form of $w(z)$, we can calculate the
improvement to the best fit,
\begin{equation}
\chi^2_{\rm eff}=-\Delta(2\ln \mathcal{L})=2\ln
\mathcal{L}(w=-1)-2\ln \mathcal{L}(w=w_i),
\end{equation}
where the likelihood function $\mathcal{L}$ is proportional to
$\exp\left(-\chi^2/2\right)$, $\mathcal{L}(w=-1)$ and
$\mathcal{L}(w=w_i)$ represent the likelihood functions calculated
in best-fit $\Lambda$CDM and the best-fit $w=w(z)$ model,
respectively. In our calculation, an improvement of $\chi^2_{\rm
eff}=4.0$ is found with $12$ additional free parameters. So the
cosmological constant can well fit the observational data. The
additional complexity in the dark energy model is not required.
\begin{figure}
\includegraphics[width=0.5\textwidth]{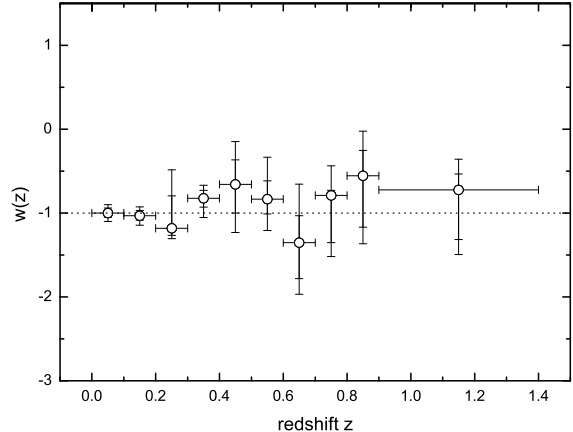}
\caption{\label{Fig3} Estimation of the uncorrelated dark energy EOS
parameters at different redshift bins ($w_1,w_2,...,w_{10}$) from
SNe Ia+BAO+WMAP9+H(z)+GRB data. The open points show the best fit
value. The error bars are $1\sigma$ and $2\sigma$ confidence levels.
The dotted line shows the cosmological constant. }
\end{figure}

\begin{figure}
\includegraphics[width=0.5\textwidth]{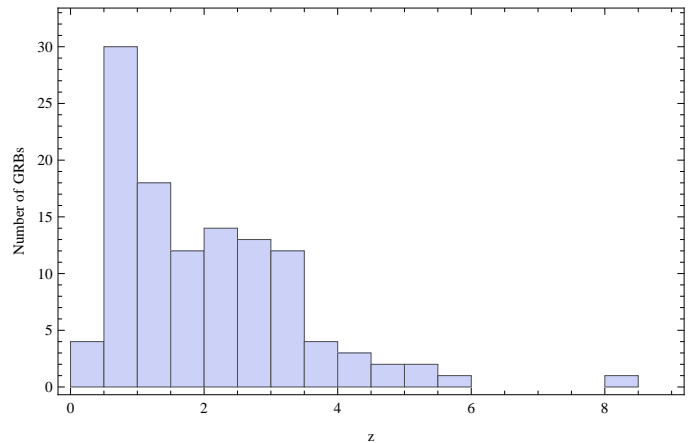}
\caption{\label{Fig4} The redshift distribution of GRB data. The
redshift range of GRBs is from $z=0.17$ to $z=8.2$, which is larger
than that of SNe Ia. The farthest SNe Ia is at $z=1.4$.}
\end{figure}

We also show the probability distributions of uncorrelated dark
energy EOS $w_i$ in Figure 2. For the first four bins, the errors
are very small, and the EOS is tightly constrained. The likelihood
distributions are close to Gaussian. At high redshift, the
probability distributions of $w_i$ span a wide range. We obtain the
$1 \sigma$ error bar around the best-fit value of the $w_i$
parameter. More standard candle data is required to reduce the
error. Long gamma-ray bursts (GRBs) are promising candidates,
because of their high luminosities and quasi-standard candle
luminosity correlations. Figure 3 shows the uncorrelated dark energy
EOS parameters at different redshift bins using SNe
Ia+BAO+Planck+H(z)+GRB data. The EOS od dark energy is almost
consistent with the cosmological constant at the $1\sigma$
confidence level. We use the GRB data from \cite{WangF}. The GRB
data covers the redshift range between $z=0.17$ and $z=8.2$. The
redshift distribution of GRB data is shown in Figure \ref{Fig4}. We
can see that almost half of GRBs are at redshift $z>1.4$, which is
the largest SNe Ia redshift in Union 2.1 sample. So long GRBs are a
promising complementary probe of dark energy at high redshifts
\cite{GRB}. Comparing Figure 1 and Figure 3, we can see that the
errors of EOS are reduced significantly by including GRBs. In order
to show the importance of observation data in the Planck era, we
also present the constraint on uncorrelated dark energy EOS
parameters from SNe Ia+BAO+WMAP9+H(z) data in Figure 5. The distance
priors from WMAP9 are used \cite{Hinshaw}, including the acoustic
scale ($l_a$), the shift parameter ($R$), and the decoupling
redshift ($z_*$). In the calculation, we use the Hubble constant
$73.8\pm 2.4$ $\rm km~s^{-1}~Mpc^{-1}$ from Cepheid variables
\cite{Riess11}. From this figure, the EOS parameter of one bin
($z\sim 0.75$) deviates the cosmological constant at the $2\sigma$
confidence level. The errors of EOS are larger than those of Figure
1. The main reason is that the distance priors from Planck are
significantly tighter than those from WMAP 9-year data
\cite{Wang13}.

\begin{figure}
\includegraphics[width=0.5\textwidth]{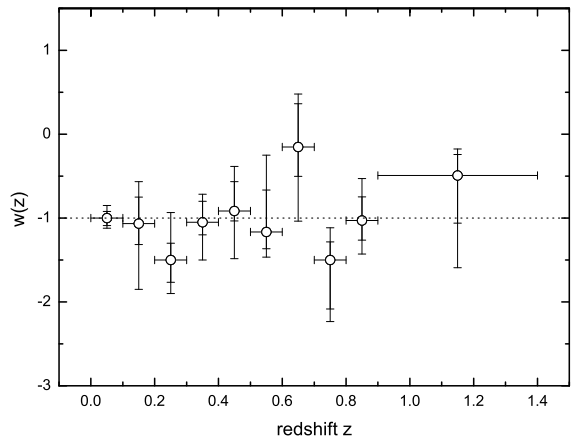}
\caption{\label{Fig5} Estimation of the uncorrelated dark energy EOS
parameters at different redshift bins ($w_1,w_2,...,w_{10}$) from
SNe Ia+BAO+WMAP9+H(z) data. The open points show the best fit value.
The error bars are $1\sigma$ and $2\sigma$ confidence levels. The
dotted line shows the cosmological constant. At one redshfit bin,
the EOS deviates the cosmological constant at the $2\sigma$
confidence level.}
\end{figure}

\section{Discussion}

Previous investigations show that the EOS of dark energy may be
consistent with the cosmological constant or prefers a dynamical
dark energy. In this paper, we apply the latest CMB from Planck, SNe
Ia, BAO from SDSS, 6dF, BOSS, WiggleZ, and the Hubble parameter
value $H(z)$ to constrain the uncorrelated EOS of dark energy using
the nonparametric technique. We model the EOS as piecewise constant
values in 10 bins. Previous studies only use less than 5 bins. We
find that the uncorrelated EOS is consistent with the cosmological
constant at the $2\sigma$ confidence level. But if we use the CMB
data from WMAP9, the results deviate from the cosmological constant
at the $2\sigma$ confidence level at one redshift bin. At present,
we find that the cosmological constant is consistent with
observations, and no preference of a dynamical dark energy. We also
find that the use of complex form of $w(z)$ does not provide a
statistically significant improvement over the use of the
cosmological constant.

With an increasing data of supernova sample, and improvement in
other cosmological observations (such as CMB, BAO, H(z)), the binned
EOS values may be constrained much more better. Future data, such as
from LSST \cite{LSST} and Euclid mission \cite{Euclid}, will shed
light on the evolution of dark energy EOS.

We thank two anonymous referees for their helpful comments and
suggestions that have helped us improve our manuscript. This work is
supported by the National Basic Research Program of China (973
Program, grant 2014CB845800) and the National Natural Science
Foundation of China (grants 11373022, 11103007, and 11033002).

\end{document}